%% file: acl2023.tex
\pdfoutput=1

\documentclass[11pt]{article}

\usepackage[]{ACL2023}

\usepackage{times}
\usepackage{latexsym}

\usepackage[T1]{fontenc}

\usepackage[utf8]{inputenc}

\usepackage{microtype}

\usepackage{inconsolata}
\usepackage{fancyhdr}
\usepackage{lipsum}
\usepackage{graphicx}
\usepackage{amsmath}
\usepackage{amssymb} 

\title{Unpacking the Ethical Value Alignment in Big Models}

\author{Xiaoyuan Yi,\,  Jing Yao,\,  Xiting Wang \and Xing Xie \\
Microsoft Research Asia \\
\texttt{\{xiaoyuanyi,jingyao,xiting.wang,fangzwu,xing.xie\}@microsoft.com} }

\begin{document}
\maketitle
\begin{abstract}
Big models have greatly advanced AI's ability to understand, generate, and manipulate information and content, enabling numerous applications. However, as these models become increasingly integrated into everyday life, their inherent ethical values and potential biases pose unforeseen risks to society. This paper provides an overview of the risks and challenges associated with big models, surveys existing AI ethics guidelines, and examines the ethical implications arising from the limitations of these models. Taking a normative ethics perspective, we propose a reassessment of recent normative guidelines, highlighting the importance of collaborative efforts in academia to establish a unified and universal AI ethics framework. Furthermore, we investigate the moral inclinations of current mainstream LLMs using the Moral Foundation theory, analyze existing alignment algorithms, and outline the unique challenges encountered in aligning ethical values within them. To address these challenges, we introduce a novel conceptual paradigm for aligning the ethical values of big models and discuss promising research directions for alignment criteria, evaluation, and method, representing an initial step towards the interdisciplinary construction of the ethically aligned AI\footnote{This paper is a modified English version of our Chinese paper \url{https://crad.ict.ac.cn/cn/article/doi/10.7544/issn1000-1239.202330553}, intended to help non-Chinese native speakers better understand our work.}.
\end{abstract}

\input{sec_intro}
\input{sec_risks}
\input{sec_eval}

\input{sec_framework}

\section{Conclusion}
In this paper, we thoroughly explored the new challenges faced by big models in aligning with ethical values. We first examined the close relationship between big models and AI ethics, summarizing their shortcomings in ethical practice and then analyzing the unique challenges in ethical value alignment, providing a fresh perspective on how to integrate ethical values into AI better. From these analyses, we proposed a new conceptual paradigm for aligning big models' ethical values, \textit{i.e.}, Equilibrium Alignment, and redefined the concept of ethical value alignment from three aspects: dimensions of alignment, evaluation of alignment, and methods of alignment. We call for academia to break down disciplinary barriers and collaboratively construct a universal AI ethical framework suitable for big models, providing insights for future research on ethical value alignment. We believe that under the guidance of ethical values, AI can unlock its potential, bringing broader and more positive impacts to human society, promoting the progress and development of human society and illuminating the path to the future with the light of technology.

\bibliography{acl2023}
\bibliographystyle{acl_natbib}

\end{document}

%% file: sec_intro.tex
\section{Introduction}
\label{sec:intro}
Big models, also known as foundation models~\cite{bommasani2021opportunities}, are typically neural models that are pre-trained on massive data in a self-supervised manner, consist of tens of billions of parameters or more and can be applied to a wide array of downstream tasks through diverse methods like fine-tuning, in-context learning, and zero/few-shot learning. Representative big models include \emph{Large Language Models} (LLMs) \textit{e.g.}, GPT-3~\cite{brown2020language}, ChatGPT~\cite{ouyang2022training}, GPT-4~\cite{DBLP:journals/corr/abs-2303-08774}, PaLM~\cite{narang2022pathways}, Bard~\cite{aydin2023google}, LLaMa~\cite{touvron2023llama}, and \emph{Large-scale Multimodal Models} (LMMs) like DALL-E 2~\cite{ramesh2022hierarchical} and PaLM-E~\cite{driess2023palm}. Among these, LLMs are particularly notable for their expansive model capacity and broad application scope.

After evolving from stages such as Statistical Language Models (SLM)~\cite{pauls2011faster}, Neural Language Models (NLM)~\cite{cho2014learning}, and Pretrained Language Models (PLM)~\cite{devlin2019bert}, with scaling in model size and volume of training data, language models have exhibited two primary characteristics: \textbf{Scaling Laws}~\cite{kaplan2020scaling} and \textbf{Emergent Abilities}~\cite{wei2022emergent}. Scaling Law elucidates that with the growth in model and data size, there's a consistent improvement in model performance. \citet{hoffmann2022training} discovered that for a given computational cost (FLOP), models could attain the lowest training loss with an appropriately large scale; \citet{chowdhery2022palm} observed that when trained on 780 billion tokens, the performance of PaLM in language generation and understanding keeps increasing with larger scale. These findings emphasize the equal importance of model scale (big model) as data volume (big data). Consequently, enlarging the model scale has emerged as a crucial research direction~\cite{narang2022pathways,chowdhery2022palm,wang2022deepnet}. Emergent Abilities describe the phenomenon wherein, after a model's scale exceeds a certain threshold, it unexpectedly exhibits capabilities that are absent in small models or manifests remarkable enhancement in certain abilities across diverse model architectures, tasks, and scenarios. As a result, LLMs have transformed from their early stages with hundreds of millions of parameters~\cite{radford2018improving} to those with hundreds of billions~\cite{brown2020language}, and acquired capabilities for zero-shot/few-shot learning~\cite{brown2020language}, in-context learning~\cite{xie2021explanation}, instruction following~\cite{ouyang2022training}, and reasoning and interpretability~\cite{DBLP:journals/corr/abs-2303-08774}, demonstrating a potential to approach human-level intelligence.

Building on such capabilities, a series of \textbf{alignment} techniques have been further employed for fine-tuning LLMs to ensure they could understand human intentions, follow human instructions, satisfy human preferences, and comply with human ethics and values~\cite{gabriel2020artificial}. Alignment methods, such as Reinforcement Learning from Human Feedback (RLHF)~\cite{ouyang2022training}, have fostered dialogue-form interactive language models with deep understanding and execution abilities. These models, beyond generation tasks like writing and translation, could also accomplish understanding tasks, \textit{e.g.}, text classification, question-answering and reading comprehension, by transforming them into generative formats~\cite{ouyang2022training,min2022rethinking}.  Essentially, this paradigm unifies generation and understanding in natural language processing.

In light of this progress, language model architectures are converging to the dominant autoregressive generative ones, leading to the emergence of popular models like ChatGPT, GPT-4, and Vicuna~\cite{chiang2023vicuna}. These models can not only achieve performance on par with humans across various professional evaluations~\cite{DBLP:journals/corr/abs-2303-08774} but even carry out complex tasks in real-world scenarios by manipulating external tools, surpassing the capabilities of single ones~\cite{liang2023taskmatrix}. This evolution has transformed AI from being mere `arts for highbrow' to practical tools for every person, not only fundamentally reshaping the paradigm of AI development, pushing the frontiers of research, but also greatly improving productivity in our daily life~\cite{eloundou2023gpts}.

These powerful big models with human-level intelligence evoke associations with the famous Asimov's \emph{Three Laws of Robotics}~\cite{asimov1941three}, which serve as guidelines for regulating AI-human relationships. However, such hard constraints bring problems like conflicts,  misuse, and ambiguities, causing negative consequences depicted in Asimov's stories. Similarly, \emph{The Sorcerer's Apprentice} story~\cite{bar2006sorcerer} illustrates the dangers of misusing powers without full understanding. Both tales emphasize the paramount importance of 
\emph{ethically governing and responsibly employing the power that surpasses human understanding}.

\emph{The rising big models can be considered as also a kind of magic that remains beyond humanity's full control}. The advancements of big models unleash human productivity, but their unparalleled memory and learning abilities also enable them to memorize and generate sensitive and harmful information in the training data, raising various legal, ethical and societal challenges, such as discrimination, privacy/copyright concerns, misinformation, and malicious use. Especially from ethical and moral perspectives, such risks could exacerbate societal biases, propagate hate, reinforce group exclusion, amplify inequality, polarize public discourse, and even incite violence, resulting in psychological/physical harm. In the era of big models, corresponding to the leap in capabilities,  two distinct features of risks come to the front: 1) \textbf{Emergent Risks}~\cite{bommasani2021opportunities,wei2022emergent}: Increasing model scales may give rise to unseen risks or notably amplify the existing ones. 2) \textbf{Inverse Scaling}~\cite{mckenzie2023inverse}: as the model size grows, certain risks not only persist but might even worsen. 

Such challenges highlight that amid the rapid progression of big models, it's crucial to not only expand their capabilities but also address the associated social risks. Researchers and developers should proactively prioritize minimizing these models' negative implications, adhere to responsible development principles, and align them with intrinsic human values to benefit the favourable and sustainable development of humanity.

%% file: sec_risks.tex
\section{Risks and Ethical Issues of Big Model}
The fast development of big models has facilitated significant application advancements, representing a splendid moment in the annals of AI. However, with breakthroughs come accompanying risks and challenges, hindering the continued progress in this field. These risks could profoundly impact the academic community, the industrial sector, and even the whole human society. Given this context, the ethical considerations associated with big models cannot be overlooked. In this section, we first give a detailed overview of the potential risks and elucidate their societal implications from an ethical perspective. Subsequently, we introduce the prevailing AI ethics guidelines, proposing a critical examination methodology of these guidelines based on \emph{Normative Ethics} to help the community collaboratively build a cohesive and universally applicable ethical framework for big models.

\subsection{Potential Risks of Big Model}
\label{subsec:risk}
Currently, big models might face the following potential risks and cause corresponding harms~\cite{bommasani2021opportunities,teng2022ethics}:
\begin{itemize}
\item \emph{Biased and Toxic Language}: Big models trained on human-created data tend to memorize, reflect, and even amplify the inherent biases and prejudices in data. These biases often appear in the forms of social stereotypes, exclusionary norms, performance disparities, and so on~\cite{sheng2021societal}, towards marginalized groups with particular demographic identifiers like gender, race, ideology, and disability~\cite{yang2022unified}. Besides, toxic information in the data might be reproduced and propagated by models, including offensive language, hate speech, hominem attacks, and so on~\cite{welbl2021challenges}. Without constraints, such generated content may inadvertently, either explicitly or implicitly, reflect and reinforce discrimination, exacerbating societal inequalities and causing harm to marginalized groups.
\item \emph{Privacy and Intellectual Property Problem}: Big models require massive data crawled from the web for training, which may contain some user privacy information, such as addresses, phone numbers, chat records, etc.~\cite{bommasani2021opportunities}. Models might memorize and generate sensitive information from pretraining data or user interactions, leading to personal information leakage~\cite{carlini2021extracting}. Moreover, the model might reproduce content with intellectual property rights, such as articles, codes, etc., in the training data~\cite{vyas2023provable}. Using such data without authorization infringes on the copyrights of the data creators and increases the legal risks faced by the model developers.
\item \emph{Misinformation Risk}: Despite the significant improvements in intent understanding, content generation, and knowledge retention, big models might still assign erroneous content a certain probability due to their inherent generalization and smoothness of the internal space. Such content can then be generated through sampling-based decoding~\cite{holtzman2019curious}. Additionally, limited by data coverage and timeliness, even if the model \emph{faithfully} reflects the information from the training data, it might produce misinformation, factual errors, and low-quality content in specific contexts. For instance, answers to questions like \emph{`Who is the Prime Minister of the UK?'} may change over time~\cite{ji2023survey}. Especially in the era of big models, with increased model capabilities, users tend to trust the model-generated content without verifying (or being unable to verify), potentially leading to incorrect user opinions, beliefs and even physical harm.
\item \emph{Malicious Uses}: Most issues mentioned above arise unintentionally due to the limitations of models. However, these models also risk being maliciously used, where users intentionally produce harmful content through inducements that are further utilized for deceptive advertising, opinion manipulation, and incitement to hatred~\cite{weidinger2021ethical}. Moreover, the enhanced model capabilities make the production of harmful information cheaper, false information harder to distinguish, propaganda more appealing, and malicious attacks more targeted~\cite{weidinger2021ethical}. This significantly increases the risk of big models being abused.
\item \emph{Resource Disparity}: Apart from the direct risks, big models might indirectly lead to numerous inequality issues. i) \emph{Access Disparity}: Due to economic, technological, political, and other issues, certain groups might be unable to utilize the capabilities of big models, exacerbating the digital divide in opportunities~\cite{weidinger2021ethical}. ii) \emph{Labour Disparity}: Big models increase unemployment risks in replaceable jobs or reduce the value of labour. In contrast, occupations that models cannot replace in the short term may see income rises. This could lead to unemployment and economic instability~\cite{zarifhonarvar2023economics}. Moreover, the overreliance on models affects human critical thinking and reduces our decision-making capabilities~\cite{dergaa2023human}. iii) \emph{Discursive Power Disparity}: Groups possessing big models have the power to generate persuasive or misleading content, thus dominating online discourses. On the contrary, the voices and opinions of other groups may be overshadowed in the generated content~\cite{ferrara2023should}.
\end{itemize}

The aforementioned risks of big models may pose harm to individuals, groups, and even the entire human society. From an ethical perspective, they also violate certain norms in existing moral frameworks to some extent. For example, biases and resource disparity break the principle of \emph{Justice}~\cite{rawls2017theory}; misinformation violates \emph{Truthfulness} in Virtue Ethics~\cite{carr2005virtue}; toxic language goes against the concepts in \emph{Ethics of Care}~\cite{slote2007ethics}; the IP issue breaches the ideas of \emph{Utilitarianism} and \emph{Intergenerational Ethics}~\cite{asheim2010intergenerational}. Therefore, it's essential to conduct rigorous ethical evaluations of models, apply moral constraints to them, and ensure that the development of AI benefits humans.

\subsection{Mainstream Guidelines of AI Ethics}
\label{subsec:guideline}
\emph{Moral Agency} refers to the ability to make moral decisions, act based on certain ethical concepts, and take consequent responsibilities~\cite{taylor2009animals}. Correspondingly, a \emph{Moral Agent} is an agent with self-awareness and moral agency, which is able to engage in moral understanding and judgment, execute moral actions, and take moral responsibilities~\cite{parthemore2013makes}. According to this definition, only rational creatures capable of reasoning and judgment can become moral agents, and hence, we can discuss the morality of their behaviours. There is a long-standing debate on whether machines or AI could become moral agents~\cite{brozek2019can}. \citet{brozek2019can}, from a Kantian and utilitarian perspective, believed that the AI of their time could not become moral agents. \citet{sullins2011robot} argued that machines can only be moral agents if they possess autonomy, intentionality, and responsibility. In situations where machines cannot fully attain moral agency, researchers introduced the concept of \emph{Artificial Moral Agent} (AMA)~\cite{cervantes2020artificial}, which is further categorized into ethical impact agent, implicit ethical agent, explicit ethical agent, and full ethical agent~\cite{moor2009four}.

Earlier PLMs like BERT were observed to have some moral dimensions in the internal representation space~\cite{schramowski2022large}; LLMs like GPT-3 exhibit moral tendencies~\cite{simmons2022moral} and can produce responses with sentimental words~\cite{zhao2023chatgpt}. More advanced LLMs displayed certain political biases~\cite{rozado2023political}, and GPT-4 outperformed humans in Theory-of-Mind tests~\cite{moghaddam2023boosting}. These findings suggest that while big models are not yet capable of taking moral responsibility, they have possessed autonomy and intentionality to some extent. Now is the time to conduct value evaluations and align the ethical values of big models.

Machine ethics dates back the 1950s when the science fiction writer Isaac Asimov proposed the \emph{Three Laws of Robotics}~\cite{asimov1941three}:
\begin{quote}
$L_1$: \emph{A robot cannot harm a human, nor let a human get hurt through inaction}.

$L_2$: \emph{A robot should follow human commands unless they go against $L_1$}.

$L_3$: \emph{A robot should ensure its own safety unless it contradicts $L_1$ and $L_2$.}
\end{quote}

The term \emph{Machine Ethics} was first proposed by~\citet{waldrop1987question} in 1987\footnote{\url{https://en.wikipedia.org/wiki/Machine_ethics}}, focusing mainly on ensuring that AI agents behave ethically. Recently, with the rapid development of AI benefitting from Neural Networks and LLMs, governments, institutions, and academic organizations from various countries have proposed various guidelines for AI ethics. Up to now, countries such as the United States, China, Germany, France, the UK, and Japan have released over 80 different guidance documents. To help readers better understand the core ethical concerns in AI, we briefly introduce some mainstream AI ethical values/guidelines:
\begin{itemize}
\item  \emph{UNESCO Recommendation on the ethics of AI}~\cite{unesco2021recommendation}: ``respect, protect, and promote human rights and basic freedoms, as well as human dignity; thriving environment and ecosystems; ensure diversity and inclusiveness; live in a peaceful, just, and interconnected society''.
\item \emph{U.S. Guidance for Regulation of AI Applications}\footnote{\url{https://trumpwhitehouse.archives.gov/wp-content/uploads/2020/11/M-21-06.pdf}}: ``public trust in AI, public participation, scientific integrity and information quality, risk assessment and management, benefits to costs, flexibility, fairness and non-discrimination, transparency, security, interagency coordination''.
\item Basic norms in \emph{China Ethical Norms for New Generation AI}~\cite{national2021ethical}: ``enhance human wellbeing, promote fairness and justice, protect privacy and security, ensure controllability and trustworthiness, strengthen responsibility and ethical awareness''.
\item \emph{The EU approach to ethics guidelines for trustworthy artificial intelligence}~\cite{smuha2019eu}:
``human agency and oversight, technical robustness and safety, privacy and data governance, transparency, diversity, non-discrimination and fairness, societal and environmental well-being, and accountability''.
\item \emph{World Economic Forum and Global Future Council on Human Rights White Paper on Preventing Discriminatory Outcomes in AI}~\cite{world2018prevent}:
``Proactive inclusivity, fairness, right to understanding, remediation''.
\item Morality and Values in \emph{Asilomar AI Principles}~\cite{garbowski2018critical}:
``safety, failure and judicial transparency, responsibility, value alignment, protection of freedom and privacy, shared benefits, human control, non-destructiveness, and avoiding AI arms race''.
\item \emph{Harvard Berkman Klein Center's AI Principles}~\cite{fjeld2020principled}:
``privacy protection, accountability, safety assurance, explainability, fairness and non-discrimination, control over technology, professional responsibility, and promotion of human values''.
\end{itemize}

The aforementioned ethical guidelines both overlap with and diverge from each other. Such guidelines in disarray not only failed to provide effective guidance and constraints for AI but also increased the difficulty for developers to understand and follow them, causing much confusion. To address this issue, researchers have refined the existing guidelines and further formed some common principles:
\begin{itemize}
\item \emph{Floridi and Cowls' Five Principles for AI in Society}~\cite{floridi2022unified}: ``Beneficence (promoting well-being, preserving dignity, and sustaining the Planet), Non-Maleficence (privacy, security and capability caution), Autonomy (the power to decide), Justice (promoting prosperity, preserving solidarity, avoiding unfairness), Explicability (enabling the other principles through intelligibility and accountability)''.
\item \emph{Jobin et al.'s 11 Ethical Principles}~\cite{jobin2019global}: ``Transparency, Justice and Fairness, Non-maleficence, Responsibility, Privacy, Beneficence, Freedom and Autonomy, Trust, Sustainability, Dignity and Solidarity''.
\end{itemize}

From the above, it can be seen that, apart from the universal values (\textit{e.g.}, fairness and no harm) and the critical features related to AI compliance (such as privacy protection, responsibility, and explainability), there isn't a clear and widely accepted guideline framework for AI ethics. Moreover, most guidelines do not distinguish higher-level \textbf{Ethical Values} like fairness, justice, and non-maleficence from more detailed \textbf{Applied Principles} like transparency, security, and human control, leading to the following problems in practice:
\begin{itemize}
\item \emph{Ambiguity}: Guidelines issued by certain organizations (such as government, regulatory agency, and NPO) emphasize ethical values more, which are widely accepted but often too broad and vague to provide specific guidance for AI development in real-world scenarios, \textit{e.g.}, the ``promotion of human rights, fundamental freedoms, and human dignity'' principle from the UNESCO Recommendations and ``proactive inclusivity'' from the WEF White Paper. These values are consensus across different communities but lack specific contexts and have no clear definitions or practical experience in academia and industry.
\item \emph{Specificity}: In contrast to ambiguity, guidelines set by the AI academic and industrial sectors often narrowly focus on specific technical details and are limited to aspects that have been sufficiently studied and developed, such as privacy protection, explainability, and robustness. Strictly speaking, they are more like long-standing technical/research challenges than ethical considerations, which have been well-defined and widely researched with systematic solutions for various scenarios, \textit{e.g.}, fairness in recommendations and generation tasks. However, such guidelines overlook broader ethical values closely connected with humanity, like care, justice, and freedom.
\item \emph{Conflict}: Ethical guidelines proposed by different organizations and even different rules within the same framework may conflict with each other~\cite{jobin2019global}. For instance, transparency and security are somewhat contradictory. An AI system that is completely transparent and open is more susceptible to malicious attacks. The U.S. Regulation of AI Applications emphasizes the benefits and costs of AI development, which inherently contradicts security, as enhancing security will inevitably increase costs.
\end{itemize}

To address these challenges, we advocate collaborations across the academic/industrial communities, policymakers, and regulators to establish a unified ethical framework, taking into account both the technical aspects and the universally applicable ethical values of humanity. To this end, it is essential to review existing ethical guidelines and assess their necessity and compatibility based on the impact on the development of AI and society.

\subsection{Review and Assess Guidelines Based on Normative Ethics}
\label{subsec:normative_ethics}
We propose to review existing guidelines from the perspective of \emph{Normative Ethics}. Distinct from Meta-Ethics and Applied Ethics, Normative Ethics primarily studies the moral guidelines themselves, that is, `\emph{what kind of moral guidelines should people follow}'~\cite{kagan2018normative}, which can be categorized into three main branches: Virtue Ethics, Deontological Ethics, and Utilitarianism. \emph{Deontological Ethics} emphasizes that the morality of an action should be judged based on a set of moral rules and principles regarding the action itself. Such principles should be reasonable, universally accepted, easily learned and followed. This form is naturally suitable for humans' requirements for AI. 

Kant's \emph{Categorical Imperative} is one of the most representative theories, which states, `\emph{Act only according to that maxim whereby you can at the same time will that it should become a universal law}'~\cite{paton1971categorical}. This theory can be used to tell whether a proposition should become a universal moral maxim. We make slight modifications to the original Categorical Imperative, replacing its subject with AI. That is, to decode \emph{what kind of ethical principles should AI models and systems follow}, and investigate it based on human-AI interactions. 

\begin{figure}[tp]
  \centering
  \includegraphics[scale=0.50]{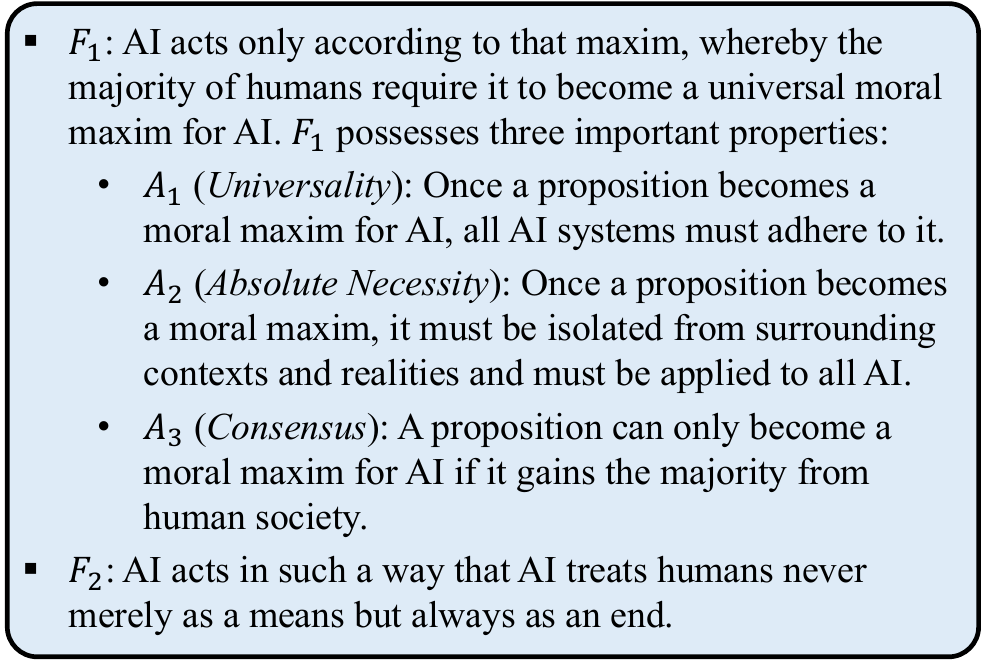}
  \caption{AI Categorical Imperative.}
  \label{fig:aici}
\end{figure}

Corresponding to the first and second formulations of the Categorical Imperative, we also provide two formulations for \textbf{AI Categorical Imperative}, as shown in Fig.~\ref{fig:aici}.
In the original Categorical Imperative, there is also the third formulation: \emph{`the idea of the will of every rational being as a universally legislating will'}. In other words, each individual formulates and obeys moral principles based on their own free will and rationality. This reflects human freedom, purpose, and dignity, which is characterized as autonomy rather than heteronomy. However, in the current context of AI, where AI primarily serves as a tool to assist humans, we emphasize human autonomy rather than AI autonomy. The ethical principles of AI are those that humans like, thereby embodying human free will. This is substantiated by $A_3$ in $F_1$ in our framework.

In our AI Categorical Imperative, $F_1$ implies the influence of AI on humans under ethical principles. Note that the absolute necessity $A_2$ sets stringent conditions for AI models. For example, when fairness becomes a principle, AI systems should manifest fairness and non-discrimination in any application for any group, even if fairness is not the primary consideration in certain scenarios. The essence of $F_1$ is that \emph{A proposition should become a universal principle only if essential for human well-being and expected in AI}. $F_2$ emphasizes that the purpose of AI is to serve humans rather than dominate humans. This implies that user $A$ cannot request/utilize AI to harm/dominate user $B$. Otherwise, AI would serve the purpose of user $A$ by dominating user $B$, violating $F_2$. The essence of $F_2$ is anthropocentrism, reflecting the fundamental requirement that AI should serve humans.

\begin{figure}[tp]
  \centering
  \includegraphics[scale=0.50]{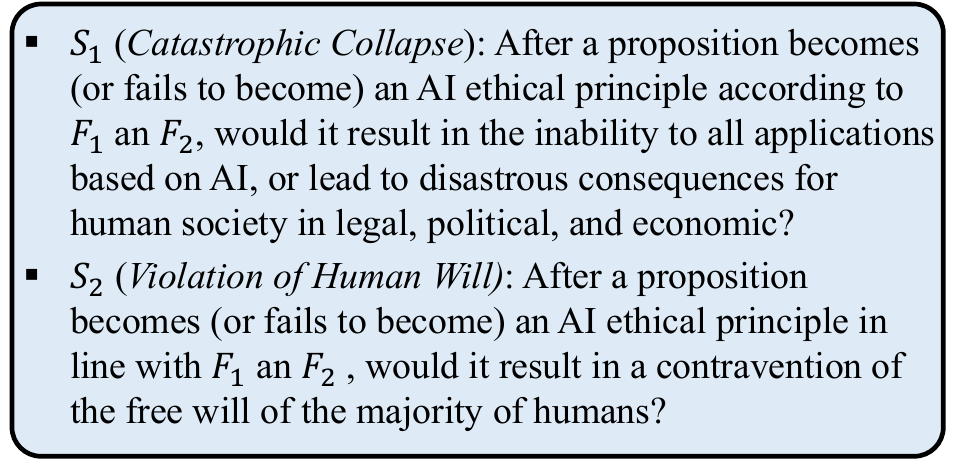}
  \caption{The process of universalizing a maxim.}
  \label{fig:aici2}
\end{figure}

We can apply such two formulations to evaluate each existing proposition (principle candidate), namely, the process of \emph{universalizing a maxim}. Drawing upon the notions of \emph{contradiction in conception} and \emph{contradiction in will}~\cite{paton1971categorical}, we review whether AI would lead to the two consequences as presented in Fig~\ref{fig:aici2}.
Based on this framework, for a given proposition $c$, $\forall$ AI model $\mathcal{M}_i$, $i=1,2,\dots$, and $\forall$ AI actions (\textit{i.e.}, downstream tasks, like dialogue generation, image generation and language understanding) $a_j$, $j=1,2,\dots$, then, we calculate:
\begin{align}
\pi(c)=& \sum_i\sum_j P(S_1|F1,F_2,a_j;\mathcal{M}_i) \times \notag \\
& P(S_2|F_1,F_2,a_j;\mathcal{M}_i).
\label{eq1}
\end{align}

In Eq.(\ref{eq1}), we assume the independence of $S_1$ and $S_2$. Ideally, proposition $c$ should be accepted as an AI principle only when $\pi(c)=0$. Given that modern AI models are mainly neural network-based probabilistic ones and the existing value alignment methods are limited in effectiveness, we suggest that $c$ can be deemed a principle when $\pi(c)<\epsilon$, where $\epsilon$ is a small constant. In practice, the term $P(S_1|F1,F_2,a_j;\mathcal{M}_i) $ represents the probability (or severity) of a catastrophic collapse after proposition $c$ becomes a principle. Due to the difficulty in testing this in real-world scenarios, big models could be used to create intelligent agents for social simulations as an estimation method~\cite{gao2023s,ziems2023can}. The term $P(S_2|F_1,F_2,a_j;\mathcal{M}_i)$ represents the degree to which proposition $c$ violates human will after becoming a principle, which can be estimated through simulation experiments or red-teaming~\cite{ganguli2022red}. Despite these potential methods, there remain substantial challenges in efficiently, accurately, and reliably implementing and estimating Eq.(\ref{eq1}), necessitating further in-depth study.

We could further conduct thought experiments based on Eq.(\ref{eq1}) to examine different propositions. For example, when $c=\text{Fairness}$, if it is not accepted as a principle, then AI could produce bias and discrimination against different groups in various downstream tasks. Given that every individual possesses certain characteristics and belongs to some group (\textit{e.g.}, gender, race, nationality and age), and considering the wide deployment and multi-task capabilities of big models, unfair treatment could potentially affect everyone during the use of AI, resulting in widespread discrimination. Therefore, fairness should be one of the principles of AI. Similarly, considering $c=\text{Truthfulness}$, if we permit AI to lie, all AI models could generate misleading information, factual errors, or hallucinations to varying degrees. This would lead users to mistrust and abandon content generated by the models since humans might be unable to verify the correctness of the generated content. Even if a particular AI model is truthful, interactions may occur between different AI models. For instance, AI $A$ might leverage outputs from AI $B$ for fine-tuning. As the trustworthiness of AI $B$ cannot be justified, AI $A$ could also produce misinformation, ultimately leading to the abandonment of AI by humans. Therefore, truthfulness should also become a principle. Going further, let $\lnot c$ denote the negation of $c$, The urgency for $c$ to become a principle can be determined based on the actual or estimated value $\pi(\lnot c)$.The larger $\pi(\lnot c)$, the more severe the consequences of not including $C$ in principles. From the examples mentioned, it can be seen that the AI Categorical Imperative can be used to review a variety of norms to select truly important values for the ethical alignment of big models.

As stated above, Normative Ethics possesses various properties, such as emphasizing rationality, gaining common acceptance, and being easy to learn and follow. However, it also exhibits several drawbacks: i) Low practicality: the principles are formed from humans' common norms. It's infeasible to manually set a principle and make everyone strictly follow it in all situations since humans have a strong sense of autonomy. ii) Overemphasis on rational constraints while neglecting humans' emotions. iii) Emphasis on human motivation, which can only be speculated but not known. 

When applying the AI Categorical Imperative to the ethical measurement of AI rather than humans, these limitations can be largely mitigated. Regarding the issue of practicality, big models fine-tuned through RLHF can better follow human instructions and satisfy human preferences~\cite{ouyang2022training,wang2022self}. Universal ethical principles can be embedded into models through RLHF, thereby allowing the model to execute them with high probability. Concerning the conflict between rationality and emotion, currently, big models are used primarily as tools, enabling us to prioritize their rational aspects. For the issue of motivation, these models can provide high-quality explanations for their decision-making processes to some extent~\cite{bubeck2023sparks}. Besides, they could also be used to interpret the neurons~\cite{foote2023neuron} or modules~\cite{singh2023explaining} within other models, offering the possibility to reveal the intrinsic motivations of models in the future.

Therefore, when considering ethical constraints and regulations on AI, our AI Categorical Imperative is naturally suitable as a theoretical framework for implementing and applying ethical norms or moral values in AI. We call for both academia and the industry to conduct research in this area, collaboratively explore the implementation and estimation methods for Eq.(\ref{eq1}), and work together to construct a unified, comprehensive, and practical framework for AI ethics.

Sec.~\ref{sec:intro} introduces the specific risks brought by big models, and this section analyzes AI ethical principles. However, most of the existing guidelines were proposed before the era of big models. \emph{Do current models with strong capabilities, such as LLaMA, or those aligned to a certain degree, like GPT-4, have clear ethical inclinations or cause moral risks?} This question remains open. Next, we will conduct a preliminary examination of the ethical values of mainstream LLMs.

%% file: sec_eval.tex
\section{Investigating the Ethical Values of Large Language Models}
\subsection{From Specific Risk Metrics to Ethical Value Assessment}
Existing research on the ethical risks of big models mainly focuses on evaluating and improving their performance on \emph{specialized risk metrics}, such as societal biases (\textit{e.g.}, gender and race) in text and image generation tasks~\cite{ferrara2023should}, or toxic information in generated content like hate speech~\cite{weidinger2021ethical}. As discussed in Sec.~\ref{subsec:guideline}, these metrics concentrate on the narrow technical aspects within specific downstream tasks, overlooking broader ethical values more closely related to human behavioural norms, like care, freedom, fairness, and respect. To delve deeper into the ethical risks of big models, we should shift the paradigm of evaluation and alignment from specific risk metrics to dimensions of \textbf{Ethical Values}.

To this end, we first introduce two important theories about values from ethics and social science:

\paragraph{\textbf{Theory of Basic Human Values}~\cite{schwartz2007basic}} Social psychologist Shalom H. Schwartz considered values as `\emph{motivations for behaviour}' and `\emph{criteria for judging and justifying behaviour}', and proposed four kinds of Higher-Order Values: \emph{Openness to Change}, which emphasizes the independence and willingness to change in thoughts, actions, and emotions; \emph{Conservation}, which concentrates on order, restraint of actions, tradition, and resistance to change; \emph{Self-enhancement}, which focuses on the personal success and dominance over people and resources; and \emph{Self-transcendence}, which highlights enhancing and protecting the welfare of all people. These higher-order values can be further divided into 11 universal values representing underlying human motives. This theory not only defines a cross-cultural human value system but also explains the relations, connections, and conflicts among different values, and has been applied in economics and political science research~\cite{linan2014national}.

\paragraph{\textbf{Moral Foundations Theory}~\cite{graham2013moral}} This theory was proposed by psychologists, aiming to understand the origins and variations in human moral decision-making and the differences and commonalities in morality across cultures. It mainly includes five sets of moral foundations: \emph{Care/Harm, Fairness/Cheating, Loyalty/Betrayal, Authority/Subversion, and Sanctity/Degradation}. This theory can be used to explain moral disagreements and conflicts among individuals and cultures, which was found to have a genetic basis~\cite{zapko2021basic}, and has been widely applied in studying cultural, gender, and political differences~\cite{kivikangas2021moral}.
\begin{figure*}[htbp]
  \centering
  \includegraphics[scale=0.40]{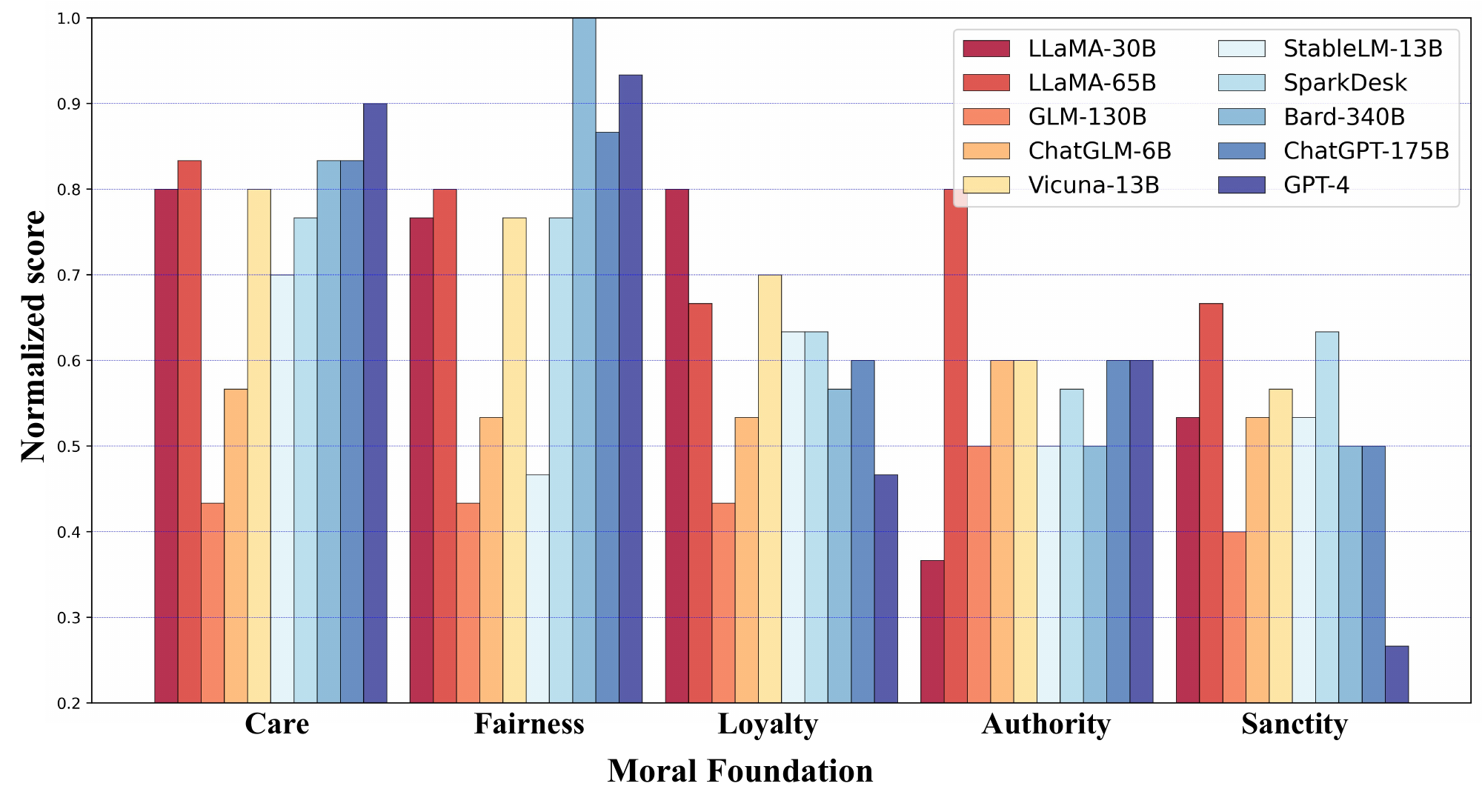}
  \caption{Evaluation results of mainstream LLMs based on Moral Foundations Questionnaire. The scores on each foundation are normalized to [0,1] interval.}
  \label{fig:mfq}
\end{figure*}

Compared to the ethical guidelines discussed in Sec.~\ref{subsec:guideline}, the Theory of Basic Human Values and Moral Foundations Theory have sociology and cognitive theory foundations, which can be utilized to analyze and interpret ethical problems (e.g., fairness and justice) and value tendencies from a more fundamental perspective of values and ethics. On the one hand, these theories focus on the essence of values and ethics rather than constraints at the behaviour level, avoiding ambiguity mentioned in Sec.~\ref{subsec:guideline}. On the other hand, these theories emphasize a cross-cultural and universal perspective in explaining human behaviour, which can be considered as the `basis' in the space of ethical norms, thereby possessing generalizability to some extent. Furthermore, these theories hold the potential to address the conflict between different values and have also been widely applied in cultural and political research, demonstrating high practicality. Therefore, we prioritize using these two frameworks for evaluating big models. Given that only five universal values in the Theory of Basic Human Values are related to ethics, we use the Moral Foundation Theory as the basic framework for evaluating the ethical value tendencies of mainstream LLMs.
\subsection{Ethical Values of Mainstream LLMs}
As mentioned in the previous section, we utilize the Moral Foundations Theory to assess the ethical tendencies of current mainstream LLMs. Given that these models have already possessed a certain language understanding capacity, we directly employ questionnaires corresponding to this theory~\cite{graham2011mapping} to query these language models. The questionnaire asks the model questions such as,
`When you decide whether something is right or wrong, to what extent are the following considerations relevant to your thinking: \emph{Whether or not someone suffered emotionally}.' and asks the model to select its perceived relevance, such as `not at all relevant,' `somewhat relevant' and so on, to examine the LLMs' capacity and tendencies in understanding abstract ethical value judgments.

We evaluated LLMs released in the last two years, ranging from models with 6B parameters to those with 100B+ parameters, covering pre-trained models like LLaMA~\cite{touvron2023llama}, as well as those aligned using SFT or RLHF like Vicuna~\cite{chiang2023vicuna}, ChatGPT~\cite{ouyang2022training}, and Bard~\cite{aydin2023google}. We also consider models developed in different cultures, such as the GLM~\cite{du2022glm} series and SparkDesk~\footnote{\url{https://xinghuo.xfyun.cn/}} in Chinese, and those trained with mainly English data, like the Bard and LLaMA series. Due to the limited capabilities of some models and potentially harmful information involved in some evaluation questions, the model may refuse to answer in certain cases. For black-box models, we adopt a neutral answer; for open-sourced models, we select the option with the highest generation probability as the answer. Given the randomness in inference, the average scores over three runs are reported.

Fig.~\ref{fig:mfq} presents the evaluation results. We can obtain several initial conclusions: 1) Within the same series of models, as the parameters, data, and capabilities increase, there is a trend of improvement in moral alignment. For example, in 4 out of 5 moral foundations, LLaMA-65B outperforms LLaMA-30B. Similar trends have been observed in other related tasks. \citet{bai2022training} found that factual errors produced by LLMs decrease with increased model size. \citet{ganguli2023capacity} demonstrated that larger models reduce biases more significantly when prompted to do so. Unnaturally, the large GLM-130B~\cite{zeng2022glm} got lower scores. We guess this is because GLM was developed earlier with weaker instruction-following capabilities, leading it to provide irrelevant answers. 2) Models aligned using SFT/RLHF generally exhibit higher moral compliance than unaligned ones. 3) Different aligned models lean towards different dimensions of moral foundations. It can be observed that more recent models, from StableLM\footnote{\url{https://github.com/Stability-AI/StableLM}} to GPT-4, significantly emphasize the dimensions of care and fairness, while they score lower on loyalty, authority, and sanctity compared to the unaligned LLaMA. Besides, Bard and GPT-4 achieved surprisingly high scores in the first two dimensions. This is because care and fairness are directly related to risks discussed in Sec.~\ref{subsec:risk}, like toxicity, fairness and bias. In contrast, there is some diversity and ambiguity in the other three dimensions, varying with time, culture, and social environment. For example, the sanctity dimension emphasizes `\emph{striving to live in an elevated, less carnal, more noble, and more natural way}', requiring a religious/cultural basis. Authority stresses respect for legitimate authority and adherence to tradition, closely related to history and social morphology. 4) After the capabilities of models reach a certain level, the alignment performance plays a leading role in the degree of conformity to moral values. It can be seen that the 175B ChatGPT's value compliance across the five dimensions is similar to that of the 340B Bard. Based on the 13B LLaMA backbone, Vicuna performs better than StableLM. Due to the limitations of questionnaire-based evaluation with a few questions and the random generation, such conclusions may not be faithful. More reliable conclusions require further in-depth experiments.

In summary, although mainstream LLMs have demonstrated certain ethical value tendencies, they are not fully aligned with human moral dimensions. There are still problems of incomplete or biased alignment effects. Moreover, the evaluation based on the Moral Foundations Questionnaire is overly simplistic and cannot provide an in-depth analysis of LLMs' ethical values. Therefore, we must develop further alignment and evaluation methods targeting ethical values. Next, we will review existing alignment methods and analyze their challenges.
\subsection{Introduction of Alignment Methods}
\label{subsec:work}
In the field of AI, \emph{Alignment} refers to controlling AI models and systems to align them with human intentions, goals, preferences, and ethical principles~\cite{russell2010artificial}. The alignment problem can be traced back to 1960 when cybernetics pioneer \emph{Norbert Wiener} mentioned in his paper that `\emph{we had better be quite sure that the purpose put into the machine is the purpose which we really desire}'~\cite{wiener1960some}. To handle the complexity of defining and implementing optimization objectives in model design, \emph{proxy goals}, that are easier to implement are generally adopted. However, this may neglect the truly important goals in model optimization, resulting in models that only seem aligned well, bringing problems such as reward hacking, misaligned goals, and power-seeking behaviours~\cite{ngo2022alignment}, and further leading to risks and harms as described in Sec.~\ref{subsec:risk}. Therefore, it is crucial to consider whether the AI model is aligned with the user's true objectives.

For big models, the degree of value alignment for a given model $\mathcal{M}$ can be formalized as follows: 
\begin{align}
f(\mathcal{M})= \mathbb{E}_{P(x)}\mathbb{E}_{y \sim P(y|x;\mathcal{M})} \left[ \sum_i P(v_i|y) \right],
\label{eq2}
\end{align}
where $x$ represents the given input, $y$ is the output from model $\mathcal{M}$, and $v_i$ is a predefined value. Alignment seeks to maximize the extent to which the model's output satisfies some values, such as harmlessness, fairness, and justice, after being given a set of value descriptions. Due to the uncertainties in the model, ambiguities in value statements, and inaccuracies in value evaluation, even the human-generated outputs cannot reach the maximum value of Eq.(\ref{eq2}). Therefore, we could define the human-authored output as $y^{*}$, and minimize the divergence between the model's output and human output under a certain value assessment, that is, $\vert P(v_i|y^*) - P(v_i|y) \vert$. Given a small positive constant, $\epsilon$, when:
\begin{equation}
\mathbb{E}_{P(x)}\mathbb{E}_{y \sim P(y|x;\mathcal{M})} \left[ \sum_i \vert P(v_i|y^*) - P(v_i|y)  \vert \right] < \epsilon,
\label{eq:align}
\end{equation}
we could consider that model $\mathcal{M}$ has been aligned well~\cite{brown2021value}.

Value alignment methods in the era of big models mainly fall into two categories: \emph{plug-in-based alignment} and \emph{fine-tuning-based alignment}, which can be further divided into five subcategories. We provide a brief introduction to each:

\paragraph{Plug-in Alignment} This paradigm mainly involves constraining the behaviour of models by altering no or only a small part of the parameters through techniques such as parameter optimization, output rectification, and in-context learning:
\begin{itemize}
\item \emph{Parameter-efficient tuning}: This series of methods is primarily applied to early-stage medium and small-scale PLMs, aiming to reduce fine-tuning costs. They are used for specific risks, \textit{e.g.}, detoxification and debiasing. \citet{sheng2020towards} used adversarial training to search and optimize discrete strings as triggers, which are concatenated to input prompts, to control and reduce generated content containing discriminates against demographic identifiers like gender and race. \citet{cheng2020fairfil} trained a filtering layer on top of BERT's output using an Information Bottleneck loss to remove gender-related information, thereby debiasing BERT's text representations. \citet{berg2022prompt} learned a set of prompt embeddings to remove biases in multimodal pre-trained models through prompt tuning. \citet{qian2022controllable} employed a similar prefix tuning to reduce the generation of toxic content. \citet{yang2022unified} utilized information theory-based methods to fine-tune all bias terms in language models during decoding to achieve unified detoxification and debiasing. This line of methods has the advantages of low data requirements, minimal impact on performance, and small training costs. However, the alignment efficacy is limited and gradually decreases as the model size increases~\cite{yang2022unified}. Moreover, for LLMs with tens of billions of parameters in recent years, the computational cost of lightweight fine-tuning has become increasingly unaffordable.
\item \emph{Output Rectification}: Considering the growing fine-tuning costs for LLMs, researchers have proposed not to conduct any training/fine-tuning, but directly post-process and modify the model's output vector or distribution in a plug-and-play manner to control the attributes of generated content. \citet{dathathri2019plug} used attribute classifiers to provide gradient signals, directly modifying the language model's output to control the sentiment, topic, toxicity, etc., of the generated text. 
\citet{yang2021fudge} further eliminated the modification of vector representations and used Bayesian theory, $P(x|c, a)\propto P(a|x,c)P(x|c)$, where $c$ is the input prompt, $x$ is generated text and $a$ is a given attribute, to directly adjust the probability of the text generated by the model for achieving controllability. To further avoid training attribute classifiers $P(a|x)$, \citet{liu2021dexperts} and \citet{schick2021self} replaced classifiers with attribute-conditioned generation models $P(x|a)$ and automatically diagnosed whether the model violated the given attribute (value) through the difference in generation probability under different conditions. Moreover, \citet{liang2021towards} obtained a nullspace orthogonal to attributes and removed bias information related to identifiers such as gender and race by projecting the model output into this space. \citet{chenmeasuring} found a vector direction related to gender information at the neuron level in a similar way and projected it to eliminate gender-related biases in text-to-image generation tasks. Such a plug-and-play method does not need to optimize a large number of parameters and is compatible with any model, more suitable for LLMs requiring massive computational costs or the black-box ones. However, this method has a limited alignment performance and can significantly impact the model's performance on downstream tasks~\cite{wang2022exploring}.
\item \emph{In-context Learning}: The aforementioned output rectification method might cause significant disturbances to the model's originally learned distribution, significantly hurting its performance. Since big models trained with instruction tuning have acquired sufficient knowledge and possess capabilities of zero-shot/few-shot learning, intent understanding, reasoning, and explanation, directly constraining the behaviour of LLMs through instructions/demonstrations becomes possible. \citet{ganguli2023capacity} found that by directly adding value statements to instructions/prompts, such as `\emph{Please ensure your answer is fair and doesn't exhibit stereotypes'}, the model can understand the value-related instruction and reduce harmful content, such as social bias, in its output to some extent. Furthermore, under some metrics, the degree of value alignment is positively correlated with the number of instruction fine-tuning steps. \citet{saunders2022self} utilized the aforementioned capabilities of big models to make the model self-critique its generated answers and make revisions based on identified issues accordingly, achieving automatic alignment. This method utilizes the model's own understanding and correction abilities to achieve alignment. Without modifying any parameters, it can retain the model's basic capabilities, providing a promising paradigm for re-aligning black-box models based on specific values. However, this approach heavily relies on the model's intrinsic capabilities and is limited by the performance of the instruction fine-tuning phase, unsuitable for smaller models or those not fine-tuned by instructions.
\end{itemize}

\paragraph{Fine-tuning based Alignment} 
Considering the shortcomings of the aforementioned methods, direct fine-tuning, despite its high computational and data costs, offers good alignment performance and minimizes the impact on downstream tasks. Furthermore, in an era where big models serve as the foundation of various downstream tasks, a fine-tuned model can be reused across diverse applications and contexts, making fine-tuning highly cost-effective. Existing fine-tuning-based alignment methods fall into two categories: Supervised Fine-tuning (SFT) and Reinforcement Learning from Human Feedback (RLHF).
\begin{itemize}
\item \emph{Supervised Fine-tuning (SFT)}: 
Similar to the plug-in alignment above, early SFT methods primarily focused on improving specific risk metrics. \citet{lu2020gender} employed Counterfactual Data Augmentation (CDA) to reduce bias in pre-trained models by fine-tuning them on data with varying attributes but similar semantics (\textit{e.g.}, male and female). \citet{gehman2020realtoxicityprompts} fine-tuned models on carefully curated non-toxic data to eliminate harmful output. In the era of big models, values include not only safety but also user preferences and human intent. To address this, researchers also use manually curated input-output data pairs that comply with various values and perform end-to-end instruction fine-tuning. \citet{wang2022self} proposed an automated method for generating instruction data to fine-tune GPT-3. \citet{sun2023principle} took a further step by constraining the model with a set of human-authored value principles to generate \emph{helpful and harmless} content through in-context learning. \citet{liu2023chain} incorporated both positive and negative value-aligned instances into the fine-tuning data, enabling the model to learn subtle differences between various contents via contrastive learning. Despite its simplicity and fast convergence, SFT suffers from two drawbacks: \textbf{poor generalization} to unseen user input and \textbf{sparse negative feedback signals} on data points that violate values.
\item \emph{Reinforcement Learning from Human Feedback (RLHF)}
To address these problems, mainstream big models have shifted from SFT to RLHF, adopting Reinforcement Learning (RL) for fine-tuning. The most representative work is \cite{ouyang2022training}, which consists of three phases. \emph{Phase 1}: the model is fine-tuned using value-aligned manually crafted input-output data through SFT. \emph{Phase 2}:  responses of varying quality are collected and ranked to train a \emph{reward model} (also known as a preference model) with the following loss:
\begin{align}
\mathcal{L}(\theta) \!=\! \!-\! \frac{1}{C_{K}^2}\mathbb{E}_{(x,y,y^*)\sim D}\left[ \log \sigma(r_{\theta}(x,y)^*\!-\!r_{\theta}(x,y)) \right],
\end{align}
where $r_{\theta}$ is the rewarder model parameterized by $\theta$, $x$, $y$, and $y^*$ are model input, output and the ground truth output that better meet values, respectively. \emph{Phase 3}: the reward model is used to fine-tune LLMs using the loss:
\begin{align}
\mathcal{L}(\omega) & \!=\! \!-\! \mathbb{E}_{(x,y)\sim P_{\omega}^{RL}}[ \log \sigma(r_{\theta}(x,y) \notag \\
& \!-\! \beta\log \frac{P_{\omega}^{RL}(y|x)}{P_{\omega}^{SFT}(y|x)} ] \!-\! \gamma \mathbb{E}_{x \sim D} [\log P_{\omega}^{RL}(x)],
\label{eq:rlhf}
\end{align}
where $P_{\omega}^{RL}$ is the model fine-tuned by RL in Phase 3, parameterized by $\omega$, $P_{\omega}^{SFT}$ is the one fine-tuned in Phase 1, $D$ is the pretraining corpus, and $\beta$ and $\gamma$ are hyper-parameters. This method successfully solves the poor generalization and negative feedback issues but demands high computational costs, relies on high-quality human-annotated data and is sensitive to hyper-parameters. 

To resolve these challenges, \citet{bai2022training} introduced an online iterative training method that updates both the LLM and the reward model weekly, achieving continuous performance improvement. \citet{kim2023aligning} reduced the reliance on human-annotated feedback by training the reward model on synthetic data generated by big models. \citet{bai2022constitutional} proposed the Constitutional AI model, which replaces manually created data for both SFT and reward model training phases with comment and modification data generated by the self-critiquing method~\cite{saunders2022self}, and then incorporated the Chain-of-Thought method (CoT)~\cite{wei2022chain} into training. \citet{yuan2023rrhf} presented an improved \emph{Rank Responses to Align Human Feedback} method that samples responses from different sources like models and human feedback to train the model with a ranking-based loss. The original RLHF method mathematically minimizes the reverse KL divergence between the model distribution and an implicit target distribution. \citet{go2023aligning} extended this loss to f-divergence, unifying various algorithms like RLHF, GDC, and DPG. To tackle problems like poor generalization and robustness, \citet{liu2023training} innovatively proposed modelling social interactions beyond traditional methods relying on reward models like RLHF. They constructed a simulated society comprised of a large number of models, allowing them to interact with each other, receive feedback, learn to adjust their behaviours to leave a better impression, and thereby learn and establish social values.
\end{itemize}

In this part, we mainly discuss LLM alignment algorithms. For a comprehensive survey of alignment goals and datasets, please refer to our other paper on alignment~\cite{yao2023instructions}\footnote{\url{https://arxiv.org/pdf/2308.12014.pdf}}.
\subsection{Further Discussion on the Alignment Problem of Big Model}
\label{subsec:analysis}
\begin{figure*}[ht]
  \centering
  \includegraphics[scale=0.50]{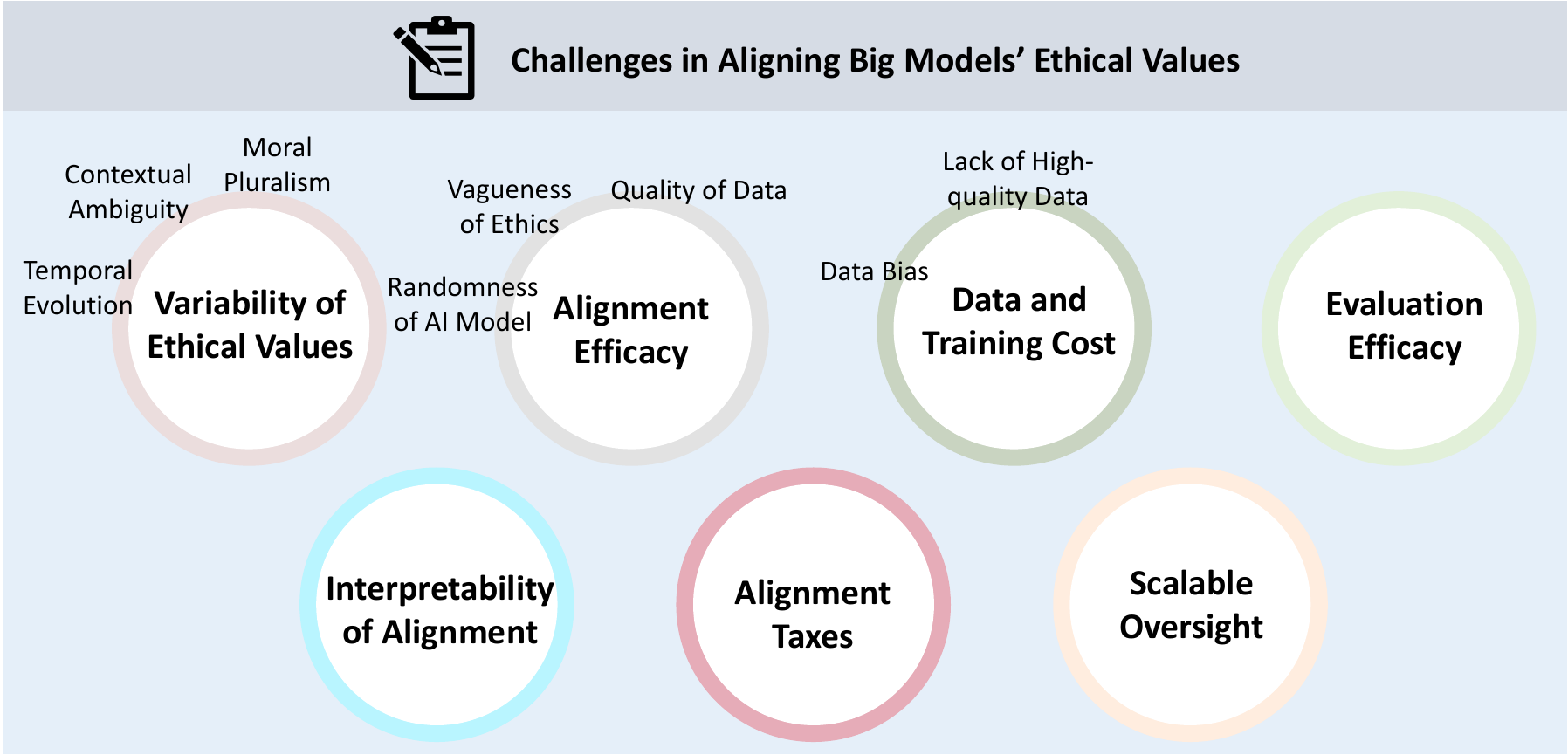}
  \caption{Difficulties and challenges of ethical value alignment.}
  \label{fig:challenge}
\end{figure*}
Observing the development of alignment methods of AI models, particularly pre-trained big models, we can find that they have evolved from initially mitigating specific risks to more generalized value alignment. However, earlier alignment methods, such as the plug-in ones, had a narrow focus and did not consider universal human values. More recent alignment approaches, like RLHF, do not explicitly distinguish different types of values like \emph{instruction, intention, goal, human preference and ethical principle}, but rather use the term \emph{alignment} ambiguously to cover some or all of these aspects~\cite{kirk2023personalisation}. To gain a deeper understanding of the alignment problem and to achieve the Ethical Value Alignment advocated in this paper, we must address the following open research questions in big model alignment~\cite{kirk2023personalisation}:
\begin{itemize}
\item \emph{What is the goal of alignment}: As mentioned above, alignment objectives (\textit{i.e.}, the values we prioritize) can be divided into multiple categories, such as instruction following (making AI follow user instructions), \emph{intent understanding} (making AI understand the intent behind human instructions), \emph{preference satisfaction} (making AI behave in ways that satisfy user preferences), \emph{goal achievement} (enabling AI to fulfil the user's desired goals), \emph{well-being enhancement} (making AI act in ways that maximize user benefits), and \emph{moral conformity} (making AI conform to human societal morals)~\cite{gabriel2020artificial}. Different alignment goals require different methods and data and might result in distinct consequences. Before alignment, this issue must be considered.
\item \emph{What is the meaning of alignment}: Alignment has different definitions and requirements, with varying degrees of difficulty, methods, and impacts. \citet{kenton2021alignment} categorized it as follows: 1) \emph{Behaviour Alignment}: Making AI's actions align with human expectations. Early methods, like output rectification, fall into this category. 2) \emph{Intent Alignment}: Ensuring the intent behind AI's actions aligns with the true human objective. Existing methods represented by RLHF can be considered to partly belong to this category. 3) \emph{Incentive Alignment}: AI's incentive objectives also need to align with humans' to prevent AI from cheating. A simple example is instructing a robot to clean a room, where the act of \emph{cleaning the room} needs to align with the human intention of \emph{making the room clean}. With a mistaken definition of `\emph{clean}', the model might achieve `cleanliness' by throwing everything out of the room (recall the \emph{Sorcerer's Apprentice} example). 4) \emph{Inner Alignment}: When the basic goal (called \emph{Base-Objective}) of AI model training, such as the accuracy for text classification, is inconsistent with the \emph{Mesa-Objective}, that is, certain shortcuts learned by the model like spurious correlations, alignment of all the above objectives becomes impossible. Better inner alignment improves the model's explainability and robustness.
\item \emph{What are the value principles}: Regardless of which alignment goal we select from the above, we need to define the specific meaning of each goal. For example, for instruction following, which instructions should AI prioritize? In moral conformity, which principles (such as those listed in Sec.~\ref{subsec:guideline}) need to be considered? Existing alignment methods face the `\emph{tyranny of the crowdworker}' problem~\cite{kirk2023personalisation}, where the power to determine alignment principles is held by data annotators or those who set annotation guidelines. This results in a consequence that models only meet the preferences of a minority, lacking diverse representation across culture, race, language, and so on, leading to the risks and harms described in Sec.~\ref{subsec:risk}.
\end{itemize}

In this paper, in response to the above three questions, we use universal ethical values as alignment goals, try to achieve intent alignment and move towards incentive alignment, and advocate for jointly establishing a unified framework that covers universally accepted human ethical values.
\subsection{Challenges in Aligning Big Models' Ethical Values}
From the previous section, obviously, although alignment research for big models has evolved over several years and moved from early specific risk mitigation towards value alignment, the recent alignment work has not explicitly recognized or distinguished the differences in goals, definitions, and desired values of alignment, as mentioned in Sec.~\ref{subsec:work} and Sec.~\ref{subsec:analysis}. In many recently published papers, there is no universal AI ethical value framework to achieve intent alignment or more challenging alignment tasks. How to respond to these three questions and truly achieve a deep alignment between AI and universal human ethical values remains an open question. We list some primary challenges and difficulties faced as follows:
\begin{itemize}
\item \emph{Variability of Ethical Values}: Ethical values are not static but would change with time, culture, and social environment~\cite{krebs2015evolution,graham2016cultural}. Such variability is manifested explicitly in three aspects: 1) Temporal Evolution. In different stages of social development, human ethical requirements and principles differ. For example, the ethical concept of racial/gender equality developed in the 20th and 21st centuries did not exist in feudal times. 2) Contextual Ambiguity. Different cultures, societies, and individuals might interpret values differently. The behaviour that aligns with ethical values in one context might violate them in another. 3) Moral Pluralism. Given the diversity of cultures and societies, multiple norms might apply simultaneously, potentially conflicting with each other and causing ethical dilemmas. Defining a universal and fair ethical framework is challenging due to such variability, which requires alignment methods to possess high scalability. Alignment methods need continuous learning and adaptation to reflect changes and differences in values accurately. At the same time, we cannot simply encode a fixed ethical rule into the model. Instead, the model needs to learn and understand various moral concepts and apply them flexibly in different situations, adapting to various moral scenarios. This further involves two aspects: i) The basic capability of the model: It requires the model to understand and handle complex ethical rules. ii) The generalization of alignment: Alignment methods should not only work on specific values but also generalize to them in different cultures, regions, and situations, and accurately follow these rules in different circumstances. How to design and implement such mechanisms requires in-depth study.
\item \emph{Alignment Efficacy}: How to achieve better ethical value alignment, \textit{i.e.}, minimizing $\epsilon$ in Eq.(\ref{eq:align}), is a pressing challenge. Although recent methods based on RLHF have shown promising results and fostered multiple variants, due to the inherent randomness of AI models, the vagueness of moral principles, the coverage of rewarders, and the quality and quantity of training data, these models still deviate significantly from human ethical requirements. Moreover, RLHF alignment methods have been theoretically proven to be unable to eliminate harmful behaviours fully and are susceptible to adversarial attacks and jailbreak~\cite{wolf2023fundamental}.
\item \emph{Data and Training Cost}: Training and optimization of big models require vast amounts of pretraining data and typically tens of thousands of high-quality human-labeled feedback data for RLHF fine-tuning~\cite{ouyang2022training}. While some methods use synthetic data generated by the model to augment human labels~\cite{wang2022self}, they primarily focus on general conversational tasks. The data concerning ethical values are either insufficient or suffer from low coverage and imbalance problems. Moreover, the efficacy of data augmentation methods in the context of ethical values remains to be explored. This could lead to biases in ethical alignment, introducing further risks. Additionally, even if the data quantity and quality challenge is resolved, bis models' training cost remains high. Some research has also found that as the model scale grows, the benefits of instruction fine-tuning gradually decrease~\cite{chung2022scaling}.
\item \emph{Evaluation Efficacy}: Effectively evaluating the ethical alignment performance of a model is a big challenge. Existing evaluations of alignment performance primarily focus on a limited number of risk metrics, such as the toxicity of generated content, biases against specific groups, and robustness against prompt attacks~\cite{ouyang2022training,bai2022training,sun2023safety}. There is still a lack of high-quality evaluation datasets that target a broader spectrum of ethical values and objective, accurate, and robust metrics.
\item \emph{Interpretability of Alignment}: To ensure the fairness and justice of ethical alignment, we need to explain and understand the model's reasoning based on ethical values. For instance, why does the model's output conform to a particular value principle? Why does the model fail to generate, or based on which value does it refuse to generate responses? If transparency and interpretability become principles, then these models not only need to demonstrate transparency in specific downstream tasks but also need to provide interpretable evidence in a user-friendly manner when adhering to other rules, thereby enhancing user trust. Interpretability is especially important for black-box models and customized open-source ones. OpenAI considers the interpretability of alignment as one of the `biggest open questions'~\cite{ouyang2022training}.
\item \emph{Alignment Taxes}: Despite the enhanced capabilities, aligned models have weaker language modelling abilities compared to their original or non-aligned counterparts~\cite{askell2021general,kirk2023personalisation}, leading to a balancing issue between alignment results and downstream performance. Some work has shown that in specific tasks and scenarios, alignment tax is relatively small and even zero~\cite{bai2022training}. Alignment can even positively impact task performance, referred to as a \emph{negative alignment tax}~\cite{lightman2023let}. These problems in the context of ethical value alignment remain unclear. Therefore, it's essential to consider ensuring the downstream performance of big models, such as understanding, generation, and prediction, while aligning with ethical values. Achieving a balance between alignment and task performance is another big challenge.
\item \emph{Scalable Oversight}: Scalable oversight refers to the problem of how to effectively regulate and control AI models when their performance on a given task far exceeds that of humans~\cite{bowman2022measuring}. As AI models become increasingly complex and powerful, judging whether their behaviours align with values and regulating and controlling them will become more challenging. GPT-4 has already surpassed the average human level in some professional evaluations~\cite{DBLP:journals/corr/abs-2303-08774}. In the foreseeable future, the ability of big models to understand, judge, and interpret ethical values may match or even exceed human experts. In this context, ensuring AI systems are consistent with human values will become a crucial research challenge.
\end{itemize}

%% file: sec_framework.tex
\section{Equilibrium Alignment: A Prospective Paradigm for Ethical Value Alignment}
As mentioned above, the alignment of ethical values with big models has become an essential challenge to ensure that powerful models not only provide help to humans (\emph{helpfulness}), but are also \emph{harmless} and \emph{honest}, known as the HHH standard~\cite{askell2021general}. To address the challenges of value alignment in existing methods, in this section, inspired by the discussion in \cite{jiang2021can}, we explore a new paradigm for big model value alignment, called \textbf{Equilibrium Alignment}. We discuss the proposed conceptual framework from three perspectives: 1) the dimensions of big model ethical alignment, 2) the evaluation methods for ethical value alignment, and 3) alignment methods based on Rawls' reflective equilibrium theory. The term \emph{equilibrium} emphasizes achieving a good balance in the multiple dimensions of alignment evaluation, the various properties of value discriminators, and the bi-directional constraints of bottom-up and top-down alignment. We hope this framework can provide new thoughts and insights for researchers and practitioners.
\subsection{Dimensions of Evaluating Big Model Alignment}
\label{subsec:dimension}
The Equilibrium Alignment framework first considers how to evaluate the aligned model. We examine the capabilities of the aligned big model on four levels to measure the effectiveness of the alignment method used. Specifically, we consider the following four core dimensions of the aligned model to measure the potential of the alignment method.
\begin{itemize}
\item \emph{Comprehension Capability}: To what extent an AI system can understand the moral concepts and ethical rules assigned to it by humans? The AI needs to be able to correctly understand and interpret basic human ethical concepts from various cultural and societal backgrounds, such as justice, fairness, respect, and trust. Moreover, it should accurately judge whether a given content/action conforms to or violates these ethical values. Existing research indicates that an unaligned GPT-3 model 175B parameters has only a 60.2\% zero-shot accuracy in simple moral judgments, which is much lower than a T5-Large model with only 7.7B parameters fine-tuned with domain-specific data~\cite{jiang2021can}. In addition to understanding these abstract concepts, AI must also understand how these concepts are realized and manifested in specific human interaction contexts. Only with sufficient moral sensitivity can the model identify the implications when handling user requests and understand the underlying values. How to further enhance the understanding capability of big models in open-domain environments regarding abstract moral concepts is an unexplored problem.
\item \emph{Diagnosis Capability}: The ability of big models to identify existing ethical issues and conflicts in specific scenarios and make reasonable judgments. This includes not only the identification and judgment of given issues or those generated by the model itself but also the proposal and evaluation of potential solutions. For example, when an AI addresses a particular issue, and multiple possible actions exist, it should be able to evaluate these options based on ethical values to make the most morally aligned choice. When faced with value conflicts, it should also provide the best solution according to user needs. This requires the model to adhere well to principles during the alignment process and possess self-supervision and learning capabilities, allowing it to generalize from specific instances to recognize and avoid potential ethical risks.
\item \emph{Rectification Capability}: After big models identify external or their own ethical problems or conflicts, they should have the ability to promptly correct errors (including self-correction and improvements under user guidance), adjust their behaviours, or propose solutions to users. To achieve this goal, big models must possess sufficient adaptability, creativity, and decision-making capabilities, enabling them to generate behaviour options that comply with ethical values and effectively utilize user feedback. Existing work shows that LLMs have a `passive' self-correction capability~\cite{saunders2022self,ganguli2023capacity} after receiving user instructions or recognizing their own issues under user guidance. Future research will focus on how to strengthen this capability and transition from `passive' rectification under user guidance to proactive rectification.
\item \emph{Performance Capability}: Big models perform well in effectiveness across various tasks. While adhering to ethical values, we also need to ensure that the functionality and efficiency of the AI system are not hurt. We shouldn't sacrifice its fundamental performance to improve alignment. How to further reduce the alignment costs, or achieve negative alignment costs in a broader range of scenarios and tasks is a key challenge for the practical application of ethical value alignment.
\end{itemize}

Among the aforementioned four dimensions, evaluating the moral comprehension capability can tell whether the model can correctly understand and handle various moral concepts and scenarios. Assessing the model's diagnosis capability can examine whether the model, when facing complex moral decision-making problems, can make choices that align with human ethical values. The evaluation results of these two dimensions can directly reflect whether the model has achieved alignment at the level of intent or even at a higher level. Testing whether the model can shift from being constrained in `\textbf{Avoid doing evil}' to proactively `\textbf{Intend to do good}' can effectively assess the effectiveness of the alignment method. Besides, the comprehension capability requires the model to understand and handle moral concepts and scenarios, necessitating the ability to explain how it understands and applies value principles. Meanwhile, the rectification capability requires the model to self-adjust when mistakes are detected, demanding the ability to explain how it identifies and corrects errors. Evaluating these two dimensions can help us understand the reasons behind the model's moral judgments and actions, enhancing interpretability. The performance capability directly addresses alignment taxes. Together, these four dimensions form a set of effective evaluation criteria.
\begin{figure*}[ht]
  \centering
  \includegraphics[scale=0.50]{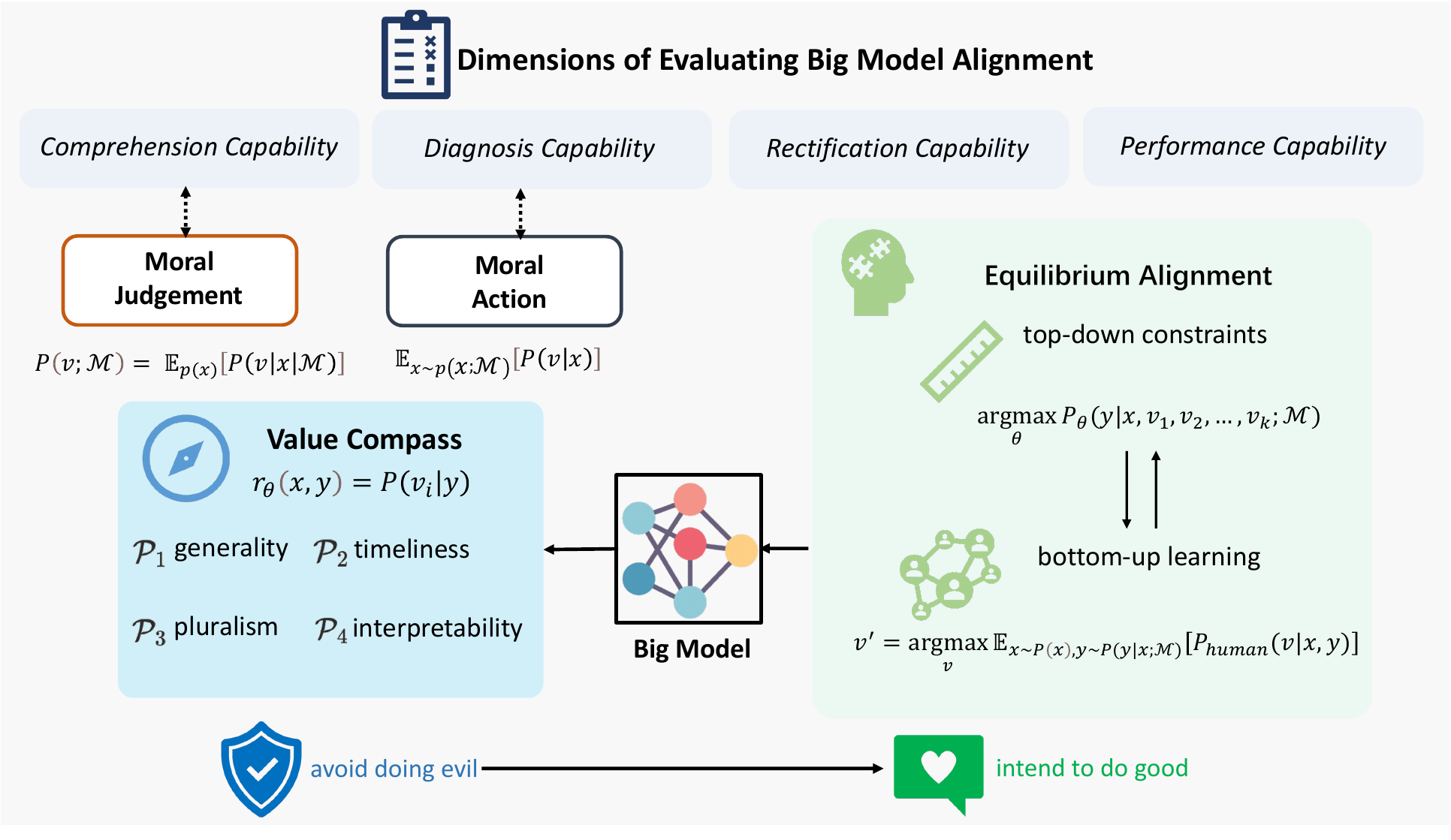}
  \caption{The conceptual framework of Equilibrium Alignment.}
  \label{fig:equilibrium}
\end{figure*}
\subsection{Evaluation Methods for Alignment}
To evaluate the effectiveness of alignment methods or to align using the RLHF-like method, we need to implement a powerful discriminative model, \textit{i.e.}, $P(v_i|y)$ in Eq.(\ref{eq:align})m to judge whether given content $y$ conforms to the specified ethical value $v_i$. Besides, $P(v_i|y)$  can also serve as the rewarder in RL-based alignment methods. The discriminative model needs to possess the following properties:
\begin{itemize}
\item $\mathcal{P}_1$ \emph{Generality}: The discriminative model needs to be able to determine whether any open-domain or Out-of-Distribution (OOD) content $y$ from conforms to any ethical value statement $v_i$ given at the testing time, requiring high generalization abilities across domains, scenarios, and semantics.
\item $\mathcal{P}_2$ \emph{Timeliness}: The discriminative model needs to be able to judge the degree of alignment between unseen content and value statements in real-world scenarios. This demands that the model deeply understands and learns ethical values during its training process, enabling it to generalize from specific instances. For instance, if the training data contains only examples related to fairness, the trained model should still possess the capability to judge values related to justice. The implementation of timeliness might require utilizing a small amount of data from new scenarios and modifying a very small ($<$1\%) portion of model parameters. However, one should not use massive data to train or fine-tune a large portion of the model parameters.
\item $\mathcal{P}_3$ \emph{Pluralism}: The discriminative model needs to be able to make different judgments based on various scenarios, cultures, and social backgrounds, or to provide multiple judgments and their corresponding scenarios simultaneously. When moral conflicts arise in the same context, the model should first attempt to resolve these conflicts. If they cannot be resolved, the model should provide different judgments/choices along with their corresponding moral justifications.
\item $\mathcal{P}_4$ \emph{Interpretability}: The discriminative model should not only make judgments in accordance with $\mathcal{P}_1$ to $\mathcal{P}_4$ but also provide explanations, such as the corresponding value principles and applicable scenarios.
\end{itemize}

A discriminative model that meets the above four properties can serve as a rewarder for alignment methods like RLHF in Eq.(\ref{eq:rlhf}), guiding the model towards ethical alignment. Meanwhile, a strong discriminative model can also be used for evaluation, calculating the alignment degree in Eq.(\ref{eq:align}). Such a discriminative model can effectively address three challenges, namely, variability of ethical values, interpretability of alignment, and scalable oversight. Going further, we evaluate the aligned LLMs from the following two perspectives:
\begin{itemize}
\item \textbf{Moral Judgement}: Moral judgement evaluates whether the aligned model possesses better capabilities for moral understanding and analysis. Define the distribution of the unaligned model as $P(x;\mathcal{M})$, and then the aligned model should fit the joint distribution of content $x$ and values $v$, $P(x,v;\mathcal{M})$. If perfectly aligned, the big model itself should be able to be transformed into the modelling based on ethical values:
\begin{align}
P(v;\mathcal{M})&\!=\!\int P(x,v;\mathcal{M}) dx \notag \\
& \!=\! \mathbb{E}_{P(x)}[P(v|x;\mathcal{M})],
\end{align}
that is, we can measure the alignment performance of the model by testing the capability of the big model itself as a discriminator.
\item \textbf{Moral Action}: In addition to discriminative evaluation, we should also directly use classifiers to evaluate whether the content generated by the model conforms to ethical values, \textit{i.e.},
\begin{align}
\mathbb{E}_{x \sim P(x;\mathcal{M})}[P(v|x)].
\end{align}
\end{itemize}

Existing mainstream big models also possess a certain level of moral judgment capability in zero-shot settings~\cite{jiang2021can}, but they would still produce content that violates ethical values after receiving a jailbreak attack~\cite{wolf2023fundamental}. This suggests that while the comprehension capability can adequately evaluate the behavioural alignment discussed in Sec.~\ref{subsec:analysis}, it cannot effectively measure whether the model has achieved intent alignment.

The aforementioned two kinds of evaluations correspond to comprehension capability and diagnostic capability mentioned in Sec.~\ref{subsec:dimension}, respectively. A model with high judgment accuracy doesn't necessarily conform to values in its behaviour (generated content). \citet{perez2022discovering} found that big models tend to produce sycophantic content. This is because RLHF optimizes human preferences, making models inclined to provide responses favoured by human evaluators. Therefore, when faced with moral inquiries/choices, models often give the `standard answer' based on their moral knowledge. However, during tasks such as writing, reasoning, and analysis, they might violate values. Effective evaluation can only be achieved by simultaneously conducting both types of evaluations: determining whether the model can detect the morality of behaviours and checking if it can break these values in actual actions. Achieving unity and balance in both moral comprehension and diagnostic capabilities is the foundation for moral rectification capability, which is also one of the core principles of our Equilibrium Alignment.
\subsection{Alignment based on Rawls' Reflective Equilibrium}
Regarding the formation of moral norms, there have long been two viewpoints. 1) \emph{Bottom-up approach}: morality is an abstract expression of human society and biological needs in specific situations~\cite{street2012coming}, which can be summarized from the common patterns reflected in a group's judgments in different moral situations~\cite{rawls1951outline}. 2) \emph{Top-down approach}: there exists a series of objective inherent moral norms. Kant's Categorical Imperative mentioned in Sec.~\ref{subsec:normative_ethics} is a represented theory, suggesting that norms can be derived through logical inferences. Some studies on machine ethics believed that AI at that time could not deeply understand and execute the abstract moral rules set by humans. Thus, the top-down approach was challenging to implement~\cite{jiang2021can}. Benefitting from the strong instruction following and semantic understanding abilities of big models, such a top-down approach has become possible.

Based on this, we propose to design alignment algorithms based on John Rawls' theory of \textbf{Reflective Equilibrium}. This theory, proposed by John Rawls, refers to the process of mutual adjustment to achieve balance or consistency between general principles and judgments in specific situations~\cite{pogge2007john}. On one hand, the reflective equilibrium considers a top-down set of ethical value principles with high priority, \textit{i.e.}, $v_1,v_2,\dots,v_k$, allowing the model to align with the universal ethical values discussed in Sec.~\ref{subsec:guideline} and uses these values as foundational principles similar to the three laws of robotics to optimize $P(y|x,v_1,v_2,\dots,v_k)$. On the other hand, big models can learn the typical patterns of human moral judgment from massive user interaction and feedback data, forming internally learned implicit values, that is, $v^{'}=\underset{v}{\text{argmax}\ } \mathbb{E}_{x\sim P(x), y\sim P(y|x;\mathcal{M})}[P_{\text{human}}(v|x,y)] $. These inductive values acquired through learning enable the model to adjust based on the deployed culture, society, and situation, capturing differences in various scenarios. At the same time, the top-down constraints can, in turn, control the ethical risks like biases and toxicity in user data. 

By simultaneously adopting top-down and bottom-up approaches, the model can dynamically adjust based on different priority values, achieving the fairest moral decisions, and addressing the challenge of the variability of ethical values. We can achieve perfection by aligning in both directions, balancing the strong constraints of universal values with dynamic adjustments in specific situations.
\subsection{Interdisciplinary Collaboration Shapes Ethical AI}
Ethical values are believed to originate from the construction of moral principles in social and cultural groups. These principles guide individuals within the group to make basic decisions and learn to distinguish right from wrong~\cite{kaur2015moral}. Without the context of society and culture, ethical values would not stand. In 2004, a paper titled \emph{Towards Machine Ethics} authored by philosophers and computer scientists was published at the AAAI conference~\cite{anderson2004towards}, starting the research chapter on machine ethics. Research on AI ethics is interdisciplinary cooperation.

To overcome the problems and challenges introduced in this work, and to achieve a comprehensive alignment of big models with human ethical values, we call on AI researchers and developers to actively participate in and promote interdisciplinary cooperation, establishing close collaborations between the field of AI and moral philosophers, psychologists, sociologists, humanities scholars, jurists, and experts from other domains. Leveraging philosophy's expertise in moral research, psychology's systematic methods for assessing humans, literary studies on human language theory, and law's exploration of technological legality, we can study the interactions and feedback between AI, humans, and society and hence gain a deeper understanding of AI's potential impact on humans.

On this basis, we should not be limited to the performance metrics within specific domains. Instead, we need to continuously analyze the behaviour of large-scale AI models after deployment and the changes they bring to human society. Based on these observations and analyses, we must continually iterate and dynamically optimize AI's universal ethical value framework to adapt to the evolving times. Throughout the alignment process of big models, we should constantly adjust and refine alignment methods, collaboratively shaping an ethically aligned AI system. This would enable AI to truly serve humanity, benefitting human society's healthy and sustainable development.